\documentclass[letterpaper, 10pt, conference]{ieeeconf}    
\IEEEoverridecommandlockouts                             
\usepackage{graphicx,float,subfigure}
\usepackage{url}
\usepackage{amssymb,amsmath,amsfonts,layout}
\usepackage{makeidx}
\usepackage{blkarray,multirow}
\usepackage{cite}
\usepackage{enumerate}
\usepackage{algpseudocode}
\usepackage{algorithm}
\usepackage{braket}
\usepackage[dvipsnames]{xcolor}
\usepackage{caption}
\usepackage[utf8]{inputenc}
\usepackage[keeplastbox]{flushend}
\renewcommand{\baselinestretch}{1}

\usepackage{tikz}
\usepackage{lipsum}
\usepackage{optidef}

\usepackage{ifthen,calc}
\usepackage{xstring}

\usetikzlibrary {positioning}
\usepackage{empheq}
\usepackage[most]{tcolorbox}

\newtcbox{\mybox}[1][]{%
    nobeforeafter, math upper, tcbox raise base,
    enhanced, colframe=blue!30!black,
    colback=blue!30, boxrule=1pt,
    #1}

\algnewcommand{\Inputs}[1]{%
	\State \textbf{Inputs:}
	\Statex \hspace*{\algorithmicindent}\parbox[t]{.8\linewidth}{\raggedright #1}
}
\algnewcommand{\Initialize}[1]{%
	\State \textbf{Initialize:}
	\Statex \hspace*{\algorithmicindent}\parbox[t]{.8\linewidth}{\raggedright #1}
}
\algnewcommand{\Outputs}[1]{%
	\State \textbf{Outputs:}
	\Statex \hspace*{\algorithmicindent}\parbox[t]{.8\linewidth}{\raggedright #1}
}

\usepackage{tikz}
\usepackage{tikz}
\usepackage{mhchem}
\usetikzlibrary{calc,arrows,automata,positioning,shapes}
\usepackage{pgfplots}
\pgfplotsset{compat=newest} 
\pgfplotsset{plot coordinates/math parser=false} 
\usepgfplotslibrary{patchplots}
\usetikzlibrary{automata,positioning}

\newcommand{\cN}{{\cal N}}

\newcommand{\cG}{{\cal G}}

\newcommand{\cC}{{\cal C}}

\newtheorem{assumption}{Assumption}
\newtheorem{theorem}{Theorem}[section]
\newtheorem{proposition}[theorem]{Proposition}

\newtheorem{definition}[theorem]{Definition}

\allowdisplaybreaks

\newcommand{\oprocendsymbol}{\hbox{$\square$}}
\newcommand{\oprocend}{\relax\ifmmode\else\unskip\hfill\fi\oprocendsymbol}

\makeatletter
\newcommand{\pushright}[1]{\ifmeasuring@#1\else\omit\hfill$\displaystyle#1$\fi\ignorespaces}
\newcommand{\pushleft}[1]{\ifmeasuring@#1\else\omit$\displaystyle#1$\hfill\fi\ignorespaces}
\renewcommand*\env@matrix[1][*\c@MaxMatrixCols c]{%
	\hskip -\arraycolsep
	\let\@ifnextchar\new@ifnextchar
	\array{#1}}
\makeatother

\DeclareMathOperator{\R}{\mathbb{R}}
\DeclareMathOperator{\C}{\mathbb{C}}

\newcommand{\diag}{\operatorname{diag}}

\newcommand{\mc}{\ensuremath{\mathcal}}

\newcommand{\brho}{\boldsymbol \rho}
\newcommand{\bPi}{\mathbf \Pi}
\newcommand{\bM}{\mathbf M}
\newcommand{\one}{\mathbf 1}

\graphicspath{{epsfiles/}}

\title{\LARGE \bf
A tractable formulation for multi-period linearized optimal power flow in presence of thermostatically controlled loads
\thanks{
E. Benenati is with ETH Zurich; email: bemilio@ethz.ch. Marcello Colombino is with the National Renewable Energy Laboratory (NREL); email: marcello.colombino@nrel.gov. E. Dall'Anese is with the University of Colorado Boulder; email: emiliano.dallanese@colorado.edu. This work was co-authored in part by the National Renewable Energy Laboratory, operated by Alliance for Sustainable Energy, LLC, for the U.S. Department of Energy (DOE) under Contract No. DE-AC36-08GO28308. Funding for M. Colombino provided by the DOE Office of Electricity, Grid Modernization Lab Consortium. The views expressed in the article do not necessarily represent the views of the DOE or the U.S. Government. The publisher, by accepting the article for publication, acknowledges that the U.S. Government retains a nonexclusive, paid-up, irrevocable, worldwide license to publish or reproduce the published form of this work, or allow others to do so, for U.S. Government purposes.
}}

\author{{Emilio Benenati, Marcello Colombino, Emiliano Dall'Anese}
}

\begin{document}

\onecolumn
\vspace{100mm}
{\Large IEEE Copyright notice}
\vspace{10mm}

© 2019 IEEE. Personal use of this material is permitted. Permission from IEEE must be obtained for all other uses, in any current or future media, including reprinting/republishing this material for advertising or promotional purposes, creating new collective works, for resale or redistribution to servers or lists, or reuse of any copyrighted component of this work in other works.

\vspace{5mm}
Accepted for publication in: 

Conference for Decision and Control 2019

December 11-13 2019 Nice, France
\newpage	

\twocolumn
\maketitle
 \thispagestyle{empty}
 \pagestyle{empty}

\begin{abstract}
This paper presents a convex reformulation of a nonlinear constrained optimization problem for Markov decision processes, and applies the technical findings to optimal control problems for an ensemble of thermostatically controlled loads (TCLs). The paper further explores the formulation and solution of a (linearized) AC optimal power flow problem when one or more ensembles of TCLs are connected to a power network. In particular, a receding horizon controller is proposed, to simultaneously compute  the optimal set-points of distributed energy resources (DERs) in the grid and the optimal switching signal for the TCLs. This formulation takes into account hardware constraints of the DERs, operational constraints of the grid (e.g., voltage limits), comfort of the TCL users,  and ancillary services provision at the substation. Numerical results are provided to verify the effectiveness of the proposed methodology. 
\end{abstract}

\section{Introduction}
\label{sec.introduction}

The paper focuses on Markov decision processes (MDPs) and outlines a methodology to formulate and solve nonlinear constrained optimization problems associated with an MDP. Technical findings are applied to management of classes of thermostatically controlled loads (TCLs) in power distribution grids~\cite{Callaway11,Mathieu12,chertkov2018optimal}, which include heating, ventilation, and air conditioning (HVAC) systems, heat pumps, and electric water heaters to mention a few. In particular, this paper considers a constrained, non-convex finite horizon optimal control problem over the evolution of an MDP. We propose a nonlinear change of variables that, for a large class of constraints, leads to an equivalent convex optimization problem which can be solved with substantial computational savings. A similar strategy was considered in~\cite{Marecki13}; however, the paper provide a more intuitive alternative proof, and the proposed method is applicable to generic convex constraints. The convex reformulation of the constrained MDP problem is then applied to management of TCLs. To this end, we propose a  discretization of the Fokker-Plank equations~\cite{Malhame85} that leads to a discrete MDP model for a system of TCLs under less restrictive assumptions than previously proposed~\cite{paccagnan15}. Motivated by the recent works demonstrating the potential of TCLs to complement DERs in providing ancillary services to assist the (re)balancing of the grid~\cite{Callaway11,Mathieu12, chen2018distributed,hao2014ancillary}, it is shown how the proposed methodology allows one to embed the MDPs optimization  into a (linearized) AC optimal power flow (OPF) problem for distribution systems to coordinate the operation of the TCLs and other distributed energy resources (DERs) in a distribution grid. Overall, the AC OPF problem accounts for cost functions associated with DERs and TCLs, operational constraints of both DERs and TCLs, and network constraints. 
The solution of such a problem lends to a synergistic control of TCLs and DERs that can ensure that the network constraints are satisfied~\cite{chertkov2018optimal}.

MDP-type models for the evolution of a population of TCLs has been first proposed in \cite{Callaway09}. Linear models for the system evolution and the control action of this model have then been proposed in various works such as \cite{Mathieu13} and \cite{Liu16}. However, it could be argued that such a model of the control action is not fully coherent with the underlying probabilistic model for the system. Other authors have proposed a control action based on modifying the deadband in which the TCL does not switch, for instance \cite{Haider15} and \cite{Bashash13}.  Recently, a different control paradigm based on energy packets inspired from communication systems has been proposed in \cite{Almassalkhi17}. The model of the control action proposed in \cite{Nijs15} is similar to the one proposed in this paper, since the controller acts directly on the transition probabilities of the system. However, \cite{Nijs15} does not take into account the constraints and optimization variables of the rest of the grid and the MDP can then be solved via value iteration.

\emph{Notation:}
Let $\R$ and $\R_{\ge0}$ denote the set of real and  nonnegative real numbers, respectively. Upper-case (lower-case) boldface letters will be used for matrices (column vectors), and $(\cdot)^\top$ denotes transposition.  For a given $N \times 1$ vector $\mathbf{x} \in \mathbb{R}^N$,  $\|\mathbf{x}\|_2 := \sqrt{\mathbf{x}^\top \mathbf{x}}$,  $\diag\{\mathbf{x}\}$ is a diagonal matrix with the elements of $\mathbf{x}$ on the main diagonal, and $x_i$ denotes the $i$-th entry. When the notation $\mathbf{x}_k, k = 1, \ldots, K$ is used to index vectors in a set, $x_{k,i}$ denotes the $i$-th entry of $\mathbf{x}_k$.  Given a matrix $\mathbf{X} \in \mathbb{R}^{N\times M}$, $X_{(m,n)}$  denotes its $(m,n)$-th entry and $\|\mathbf{X}\|_2$ denotes the $\ell_2$-induced matrix norm. When the notation $\mathbf{X}_k, k = 1, \ldots, K$ is used to index matrices in a set, $X_{k,(m,n)}$ denotes the $(m,n)$-th entry of $\mathbf{X}_k$. The matrix $\mathbf{I}_n \in \R^{n\times n}$ denotes the identity matrix. The vector $\one_n$ is the vector of all ones in $\R^n$ and vector $\mathbf{0}_n$ is the vector of all zeroes in $\R^n$.  We denote with $\mathbb{P}(\cdot)$ the probability of an event. The notation $\mathbb{P}(A|B)$ denotes the probability of event $A$ conditioned on event $B$ and $\mathbb{P}(A,B)$ denotes the joint probability of events $A$ and $B$.

\section{Constrained optimization of MDPs}\label{sec.mdps}

{An MDP is a discrete time stochastic control process that allows to model situations where the outcomes are partly random and partly under the control of a decision maker \cite{White89}. In this section we consider a class of nonlinear, stochastic finite horizon optimal control problems that can be modeled in this framework and we show that, under suitable variable transformations, they can be cast as convex optimization problems and thus solved efficiently. 
}
{We denote by} $\Omega = \{1, \dots , N\}$ {the discrete \emph{state space} of the MDP.} {We want to model the evolution of a large population of agents. Let $s_i^t\in \Omega$ denote the state of agent $i$ at time $t$ with $i=1,\dots, K$ $K\gg1$. Since modeling the evolution of each agent is impractical, we will consider the evolution in time of the} probability distribution over the states at time $t$. {We denote this distribution by} $\brho^t = [\rho^t_1, \dots, \rho^t_{N}]^{\top}{\in\R^N_{\ge 0}}$. Each $\brho$ must lie in the probability simplex 
\[
\mathcal{S}^N:=\left\{\brho\in\R_{\ge 0}^N\,|\,\mathbf{1}_N^{\top}\brho =1\right\}.
\]
{The evolution of the state of each agent is stochastic and, for every time-step, it follows the law}
\begin{align} \label{eq.prob}
 \mathbb{P}(s_k^{t+1} = i | s_k^t = j) = \Pi^t_{(i,j)}~,\quad {i, j\in\Omega}
\end{align}
that is, we assume that the probability of transitioning to a certain state does not depend on the previous states of the system (Markovian assumption). {Let $\mathbf{\Pi}^t\in\R^{N\times N}$ be the \emph{transition probability matrix}, with elements $\Pi^t_{(i,j)}$ as defined in~\eqref{eq.prob}.
{Given an initial distribution $\brho^0\in\mathbf P^N$, the distribution evolves over time according to}
\[ 
 \brho^{t+1} = \bPi^{t} \,\brho^t ~, ~  t = 0,\dots,T-1.
\]
We define the set of \textit{valid} transition probability matrices as}
\begin{align}\label{eq.transition.matrix.set}
\mc P^N :=  \left\{ \bPi\in\R_{\ge0}^{N\times N} ~ \left |~  \one_N^\top \bPi = \one_N^\top\right. \right\}.
\end{align}
The matrices in this set are \textit{valid} in the sense that, if $\bPi^{t}\in\mc P^N $, then $\brho^t\in \mathcal{S}^N,~ t = 1,\dots,T$.
{In the following, we assume that some of the transition probabilities are controllable, that is, they can be modified by the decision maker in order to minimize a cost function while satisfying certain constraints. }
{
\subsection{Markov Decision Process}
}
{Suppose the decision maker is allowed to} modify certain {transition probabilities and, given an initial probability distribution $\bar\brho$, wishes to solve the following finite horizon optimal control problem
\begin{align}\label{eq.opt.basic}
\begin{split}
\min_{\bPi^t\in\mc P^N,\brho^t\in\R^N_{\ge0}}~&~ f(\brho^1,\dots,\brho^T)\\
\text{subject to:}~&\brho^{t+1} = \bPi^t \brho^t,\quad~ t=0,\dots,T-1\\
~&~ \mathbf{c}(\brho ^0,\dots,\brho ^T,\bPi ^0,\dots,\bPi ^{T-1}) \le \boldsymbol{0}_\nu,\\
~&~\brho^0 = \bar\brho ,\\
\end{split}
\end{align}
where $f:\R^{NT}\to\R$ is a generic cost function formulated on the probability distributions and $\mathbf{c}$ is a constraint function, vector valued with dimention $\nu$ . To put formulation \eqref{eq.opt.basic} into perspective, the MDP can be thought to reach an absorbing termination state after $T$ timesteps with probability $1$. This formulation can therefore remind a Stochastic Shortest Path (SSP) problem (as defined in \cite{BertsekasDynamicProgramming}). However, the cost of an SSP problem is a function of the visited states and actions taken along the path, while in formulation \eqref{eq.opt.basic} the objective can be any generic function of the probability distribution. Clearly problem~\eqref{eq.opt.basic} is in general intractable. This is because of the bilinear terms in $\bPi$ and $\brho$ and the generic non-convex cost function and constraint set. In the following, we define classes of \emph{tractable} functions $\mathbf{c}$ that will allow for a convex reformulation of~\eqref{eq.opt.basic}.  
}
{
\begin{definition}[Tractable constraints]\label{def.tractable}
We call a constraints function \mbox{$\bf{c}:\R^{NT}\times\R^{NT\times N}\to\R^\nu$} a \underline{tractable constraint} if every element $c_i,\, i = 1, ..., \nu$ of  $\mathbf{c}(\brho^0,\dots,\brho^T,\bPi^0,\dots,\bPi^{T-1})$ is of one of the following forms
\begin{subequations}\label{eq:problem:convexificable_constraints}
\begin{align}
 &c_i(\brho^{0},...,\brho^T),\, c_i~\text{convex},  \label{eq.g1}\\
 \begin{split}
 &c_i(\bPi^0\diag(\brho^{0}),...,\bPi^{T-1}\diag(\brho^{T-1}), \brho^{0},...,\brho^T),\\
 &\hspace{0.7\columnwidth} c_i~\text{convex}, \label{eq.g2}
 \end{split}\\
 &\sum_{i=1}^N \alpha_i \Pi_{(i,j)}^t - \beta,~ t\in\{0,\dots,T-1\}.\label{eq.g3}
\end{align}\end{subequations}
\end{definition}
Note that in Problem \eqref{eq.opt.basic} every element of $\bPi$ is controllable, but the set of controllable transitions can be restricted by imposing constraints of the kind $\Pi_{(i,j)}^t = \beta$, which are of the form \eqref{eq.g3}.  We now require the following assumptions to hold:
\begin{assumption}\label{ass.tractable}
The function $f:\R^{NT}\to\R$ is convex and $\mathbf{c}:\R^{NT}\times\R^{NT\times N}\to\R^\nu$ is tractable (as in Definition~\ref{def.tractable}).
\end{assumption}
\begin{assumption}\label{ass.feasible} Let $\mathbf{c}:\R^{NT}\times\R^{NT\times N}\to\R^\nu$ be tractable (as in Definition~\ref{def.tractable}) and $\mc I :=\{i\in\{1,...,\nu\}\,|\,c_i~\text{is of the form~\eqref{eq.g3}}\}$. There exist at least one set ($\bPi^0, ..., \bPi^{T-1}$) such that the equations
\begin{subequations}\label{eq.feasibility}
\begin{align}
\bPi^t&\in\mc P^N, &t=0,\dots,T-1\\
c_i(\cdots,\bPi^0,\dots,\bPi^{T-1})&\le 0,& i\in\mc I
\end{align}
\end{subequations}
are satisfied.
\end{assumption}
Note that, since $\forall i\in\mc I$ the functions $c_i$ do not depend on $\brho^t$ and are linear in $\bPi^t$,~\eqref{eq.feasibility} is a linear feasibility problem and Assumption~\ref{ass.feasible} can be easily tested. Furthermore, Assumption~\ref{ass.feasible} simply states that the ``input constraints" on the decision variables $\bPi^t$ are not inconsistent with each other and is therefore very reasonable. 
\begin{proposition}[Convex reformulation]\label{propo.convex} Under Assumptions~\ref{ass.tractable} and~\ref{ass.feasible}, with the nonlinear change of variables \begin{align}\label{eq.change.variables}
\bM^t : = \bPi^t\diag(\brho^{t}),\quad t=0,\dots,T
\end{align} 
the optimization problem~\eqref{eq.opt.basic} is equivalent to the convex optimization problem 
\begin{align}\label{eq.opt.mod}
\begin{split}
\min_{(\brho^t,\bM^t)\in\mc M^N}~&~ f(\brho^1,\dots,\brho^T)\\
\text{subject to:}~&\brho^{t+1} = \bM^t\one_N,\quad~ t=0,\dots,T-1\\
~&~\tilde{\mathbf{c}}(\brho^0,\dots,\brho^T,\bM^0,\dots,\bM^{T-1}) \le \boldsymbol{0}_\nu,\\
~&~\brho^0 = \bar\brho\\
\end{split}
\end{align}
where every $i^{\text{th}}$ element of $\tilde{\mathbf{c}}$ is convex and defined as
\begin{align*}
& \tilde{c}_i := c_i(\brho^{1},...,\brho^T),\, \text{ if } c_i \text{ is of the form}~\eqref{eq.g1}\\
& \tilde{c}_i := c_i(\bM^0,...,\bM^{T-1},\brho^{1},..,\brho^T),\,\text{if } c_i \text{ is of the form}~\eqref{eq.g2}\\
& \tilde{c}_i := \sum_{i=1}^N \alpha_i M_{(i,j)}^t - \beta\rho^t_j,~\text{ if } c_i \text{ is of the form}~\eqref{eq.g3}
\end{align*}
and the set $\mc M^N$ is defined as
\[
\mc M^N =\left\{(\brho, \bM)\in\R^N_{\ge0}\times \R_{\ge0}^{N\times N}\, \big| \one_N^\top \bM = \brho^\top  \right\}.
\]
Furthermore, given an optimal solution $\brho^{t}_\star,~t=0,\dots,T$ and $\bM^t_\star,~t=0,\dots,T-1$, the optimal matrices $\bPi^t_\star,~t=0,\dots,T-1$ can be obtained as 
\begin{align}\label{eq.reconstruction}
\Pi^t_{\star (i,j)} = \left\{
\begin{array}{cc}
 \frac{M^t_{\star (i,j)}}{\rho^t_{\star j}} ~&~ \text{if }\rho^t_{\star j}>0\\
\textnormal{anything feasible} ~&~ \text{if }\rho^t_{\star j}=0 .\\
\end{array}\right.
\end{align}
\end{proposition}
\emph{Proof:} Let us denote by $f_1^\star$ the optimal value of problem~\eqref{eq.opt.basic} and  by $f_2^\star$ the optimal value of problem~\eqref{eq.opt.mod}. Clearly $f^\star_1\le f^\star_2$ as if $\brho^{t}_\star,~t=0,\dots,T$ and $\bPi^t_\star,~t=0,\dots,T-1$ are optimal for ~\eqref{eq.opt.basic}, then $\brho^{t}_\star,~t=0,\dots,T$ and $\bPi^t_\star\diag(\brho^t_\star),~t=0,\dots,T-1$ are feasible for problem~\eqref{eq.opt.mod}. Next we need to show that, given the optimal solution $\brho^{t}_\star,~t=0,\dots,T$ and $\bM^t_\star,~t=0,\dots,T-1$ of~\eqref{eq.opt.mod}, we can always reconstruct a feasible solution for~\eqref{eq.opt.basic}. To do so, consider the matrices $\bPi_\star^t$ constructed using~\eqref{eq.reconstruction}. Together with $\brho^{t}_\star,~t=0,\dots,T$, they are feasible for the constraints of the form~\eqref{eq:problem:convexificable_constraints}. To see this, note 
\[
\tilde c_i(\brho_\star^0,\dots,\brho_\star^T,\bM_\star^0,\dots,\bM_\star^{T-1}) \le 0
\]
immediately implies 
\[
c_i(\brho_\star^0,\dots,\brho_\star^T,\bPi_\star^0,\dots,\bPi_\star^{T-1}) \le 0,\\
\]
for constraints of the type~\eqref{eq.g1} and~\eqref{eq.g2}. For those of the type~\eqref{eq.g3} we note that, if  $\rho_{\star j}^t>0$,
\[
\sum_{i=1}^N \alpha_i M_{\star (i,j)}^t - \beta\rho^t_{\star j} \le 0 \implies \sum_{i=1}^N \alpha_i \Pi_{\star (i,j)}^t - \beta \le 0
\]
and if $\rho^t_{\star j}=0$, then, we choose any $\Pi_{\star (i,j)}\ge 0$ such that $\sum_{i=1}^N \Pi_{\star (i,j)}=1$ and
\[
\sum_{i=1}^N \alpha_i \Pi_{\star (i,j)}^t - \beta \le 0.
\]
At least one such selection is guaranteed to exist by Assumption~\ref{ass.feasible}. Next we show that $\boldsymbol{\Pi}_{\star}^t\in\mc P^N$. To see this, note that $\Pi_{\star (i,j)}\ge0$ and, if $\rho_{\star j}=0$, then $(\one_N^\top \bPi_{\star})_j $ = 1 by construction. If $\rho_{\star j}>0$, since $\one_N^\top  \bM_{\star} = \brho_\star$, 
\[
(\one_N^\top \bPi_{\star})_j =\frac{1}{\rho_{\star j}} \sum_{i=1}^N M_{\star (i,j)}=\frac{\rho_{\star j}}{\rho_{\star j}} = 1
\] 
and, therefore, $\boldsymbol{\Pi}_{\star}\in \mathcal{P}^N$. Finally, if we denote by $\mc J^t:=\{j \in \{1,...,N\}\,|\,\rho_{\star j} >0\}$  
\begin{align*}
\rho^{t+1}_{\star i} &= \sum_{j=1}^N M_{\star (i,j)}^{{t}}  =  \sum_{j\in\mc J^t} \Pi_{\star (i,j)}^{{t}} \rho^t_{\star j} =  \sum_{j=1}^N \Pi_{\star (i,j)}^{t}\rho^t_{\star j}.
\end{align*}
Since $\brho_\star^0,\dots,\brho_\star^T,\bPi_\star^0,\dots,\bPi_\star^{T-1}$ are feasible for~\eqref{eq.opt.basic}, $f^\star_2\le f^\star_1$ and the proof is complete. \hfill $\Box$
}
Not{e} that, from \eqref{eq.change.variables} and from the definitions of $\boldsymbol{\Pi}^t$ and $\brho^t$, we can write
\begin{align*}
 & M_{(i,j)}^t = \Pi^t_{(i,j)} {\rho_j^t }= \mathbb{P}(y(t+1) = i | y(t) = j) \mathbb{P}(y(t) = j) \\
 & = \mathbb{P}(y(t+1) = i, y(t) = j) \,, 
\end{align*}
which allows us to interpret the element $M^t_{(i,j)}$ as the joint probability of being in state $j$ at time $t$ and at state $i$ at time $t+1$ (instead, $\Pi^t_{(i,j)}$ is the probability of transitioning to $i$ knowing that the state at $t$ is $j$). Prop{osition}~\ref{propo.convex} {offers a simpler and more direct proof of similar results that appeared in}~\cite{Marecki13}, where {the authors propose} the solution of an MDP using a convex optimization problem over the joint transition probabilities.

\section{An alternative discretization for a population of TCLs}\label{sec.prob.definition}

{Thermostatically controlled loads (TCLs)} represent electric appliances such as HVAC systems, fridges, heat pumps and water heaters, that inherently present a thermal capacity. By exploiting this capacity to store energy, the power consumption can be deferred over short periods of time without impacting the comfort of the final user. {The control of a population of TCLs can then lead to increased flexibility in shaping the short-term energy demand and allows to compensate for some of the stochastic behavior of renewable generation.}
{The} scope of this section is to model a population of TCLs as an MDP and show that, {under basic assumptions, Proposition \eqref{propo.convex} can be applied to several finite horizon optimal control problems of interest}.
{Similarly to~\cite{paccagnan15}, we build a dynamical model starting with the Fokker Plank equation that describes the evolution of the probability density function of the TCL temperature over time, but we propose an alternative discretization of the dynamics that allows to obtain a \textit{valid} (as in \eqref{eq.transition.matrix.set}) transition probability matrix under weaker assumptions.}

\subsection{PDE modeling of TCLs} \label{sec.tcl_model}

We consider a population of {TCLs (in this example, heating systems)}.  {The evolution of the temperature} of each TCL is described by the {stochastic} differential equation
\begin{equation}
\label{eq:Model:tcl_ode}
dX = \left(-\frac{1}{CR} (X-\theta_a) + \Psi \frac{\eta P_{h}}{C}\right) d\tau + \sigma d\omega \, , 
\end{equation}
where $X$ is the temperature of the system, $R$ and $C$ are the thermal resistance and capacity, $ \theta_a $ is the external temperature, $P_{h}$ is the power consumption, $ \eta $ is a coefficient of performance and $d\omega$ is a Wiener process with variance $\sigma$, representing the uncertainties in the model. We will denote by $\tau$ the time when considered in continuous form. The system can either be ON or OFF, which is captured by the binary variable $\Psi \in \{0,1\}$. Each TCL automatically switches when the temperature exits a deadband $[\theta_-; \theta_+]$, which represents the range of temperatures that are considered comfortable for the user
\begin{equation}
\label{eq:Model:standard_control}
 {\lim_{d\tau\downarrow 0}}\Psi(\tau+d\tau) =
\begin{cases} 
	1, &	X{(\tau)} \le \theta_- \\
	0, & 	X{(\tau)} \ge \theta_+ \\
	 \Psi(\tau), &		\text{otherwise}.
\end{cases}
\end{equation}
We define then the probability density functions in the modes ON and OFF as $\mu_{\psi}(x,t) = \mathbb{P}(X = x, \Psi = \psi )|_t$. When $x\in [\theta_-; \theta_+]$, the evolution in time of these functions is given by the Fokker-Planck equations \cite{Callaway09}
\begin{equation}
\label{eq:Model:fokkerplanck}
\frac{\partial \mu_{\psi}(x,\tau)}{\partial \tau} = - \frac{\partial}{\partial x}( \mu_{\psi}(x,\tau) f_{\psi}(x) ) + \frac{\sigma^2}{2}\frac{\partial ^2 \mu_{\psi}(x, \tau)}{\partial x^2}
\end{equation} 
where $ f_{\psi}(x)$ is the r.h.s of~\eqref{eq:Model:tcl_ode}.
\subsection{{Asymmetric discretization}}

\begin{figure}
  \includegraphics[width=\linewidth]{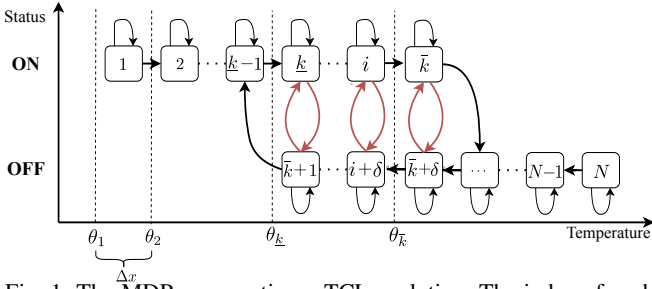}
  \caption{The MDP representing a TCL evolution. The index of each bin represents its position in the vector $\brho$. The temperature range of each bin is shown on the $x$-axis. The red arrows represent the transitions that the controller can modify. The transitions due to noise were omitted for simplicity. }
  \label{fig:TCL_diagram}
  \vspace{-2em}
\end{figure}

Let us now proceed with the discretization of the dynamics of the system. We take an uniform temperature grid with coarseness $\Delta x$ in the temperature interval $[\theta_{--}; \theta_{++}] \supset [\theta_- ; \theta_+]$, where $\theta_{++}$ and $\theta_{--}$ are temperatures that the system has a low probability to reach.
The probability of a TCL being in the temperature bin $k$ and in ON state $\psi$ at time $\tau$ is defined as
\begin{align}
\label{eq:Model:rho_definition}
    \bar{\rho}(k,\tau) := \int_{\theta_k}^{\theta_{k+1}} \mu_1(x,\tau) dx
\end{align}
By taking the derivative with respect to time of this expression and using~\eqref{eq:Model:fokkerplanck}, we obtain
\begin{align}
\label{eq:Model:finite_volume_method}
\frac{d \bar{\rho}(k, \tau)}{d\tau} = \Big[ -\mu_1(x,\tau)f_1(x) + \frac{\sigma^2}{2} \frac{\partial }{\partial x}\mu_1(x,\tau) \Big]^{x=\theta_{k+1}}_{x=\theta_k}.
\end{align}
We can obtain an equivalent expression $\underline{\rho}(k,\tau)$ for the OFF states by substituting $ \mu_0$ to $\mu_1$.
 Since $\Delta x$ is assumed small, {we approximate $\mu_{\psi}$ to be} constant within each bin. In particular, the probability density over bin $k$ (which range is $\theta_k \leq x \leq \theta_{k+1}$) is assumed to have the value of the probability density computed at the extreme temperature of the bin towards which the system is evolving, that is, $\theta_{k+1}$ if the system is heating ($\psi=1$), $\theta_k$ if it is cooling ($\psi=0$). We refer to this choice as ``asymmetric discretization". The advantages of this choice will be discussed later in this section. The resulting approximation of ~\eqref{eq:Model:rho_definition} is
\begin{align}
\label{eq:Model:approximation}
\begin{split}
&  \mu_{0}(\theta_k,\tau) \approx \frac{\underline{\rho} (k, \tau) }{\Delta x} \\
& \mu_1(\theta_{k+1},\tau) \approx \frac{ \bar{\rho} (k, \tau)} {\Delta x} .
\end{split}
\end{align}
 Notice that, from~\eqref{eq:Model:standard_control}, the probability of being ON with a temperature $ x \ge \theta_+ $ is $0$ and the same holds for temperatures $x\le \theta_-$ while being OFF. 
Let us then define the bins $\bar{k}$, ${\underline{k}}$ such that $\theta_{\bar{k} +1} = \theta_+ $ and $\theta_{\underline{k}} = \theta_-$.  We define the vector $\brho(\tau) \in \R^{N}$, 
\[
\brho(\tau) := [\underbrace{\bar{\rho}(1, \tau) \dots \bar{\rho}(\bar{k}, \tau)}_{\rho_1(\tau),\dots,\rho_{\bar k}(\tau)}, \underbrace{\underline{\rho}(\underline{k}, \tau), \dots , \underline{\rho}(n, \tau)}_{\rho_{\bar k +1}(\tau),\dots,\rho_{N}(\tau)}]^{\top}
\]
{which represents the probability distribution over all the $N$ bins that can be reached by the TCL (i.e., excluding those with zero probability).} {Substituting~\eqref{eq:Model:approximation} into~\eqref{eq:Model:finite_volume_method} we obtain a (continuous-time) linear system that approximates the natural evolution of the system as}
\begin{equation}
\label{eq:Model:linear_system_continuous}
\frac{d}{d\tau}\brho(\tau) = \mathbf{A}^{\text{nat}}\brho(\tau)
\end{equation}
The matrix $\mathbf{A}^{\text{nat}}$ is given by
\begin{equation}
\mathbf{A}^{\text{nat}}:= \frac{1}{\Delta x} \begin{bmatrix}
\mathbf{A}^{\text{on}} & \mathbf{0}_{N\times N} \\
\mathbf{0}_{N\times N} & \mathbf{A}^{\text{off}}
\end{bmatrix}  + \mathbf{A}^{\text{switch}}
\end{equation}
{Where $\mathbf{A}^{\text{on}}$, describes the time evolutions of the TCLs in the ON status,  $\mathbf{A}^{\text{off}} \in \R ^{\frac{N}{2}\times\frac{N}{2}}$ that in the OFF status and $\mathbf{A}^{\text{switch}}$ describes the transitions due to the thermostat.} The nonzero elements of the matrices $\mathbf{A}^{\text{on}}$, $\mathbf{A}^{\text{off}} \in \R ^{\frac{N}{2}\times\frac{N}{2}}$ are
\[
A^{\text{on}}_{(i, j)} = 
\begin{cases} 
\frac{\sigma^2}{\Delta x} & \text{if } i = j-1 \\
-f_1(\theta_{i+1}) -\frac{2\sigma^2}{\Delta x} & \text{if } i = j \\
f_1(\theta_{i+1}) +\frac{\sigma^2}{\Delta x} & \text{if } i = j+1
\end{cases}
\]
\[ A^{\text{off}}_{(i, j)} = 
\begin{cases} 
- f_0(\theta_i) + \frac{\sigma^2}{\Delta x} & \text{if } i = j-1 \\
f_0(\theta_{i}) -\frac{2\sigma^2}{\Delta x} & \text{if } i = j \\
\frac{\sigma^2}{\Delta x} & \text{if } i = j+1
\end{cases}
\]
For $1<i<\frac{N}{2}$. Introducing reflecting boundary conditions at the endpoints of the grid, we have
$$
A^{\text{on}}_{(i,1)} =\begin{cases}
-f_1(\theta_2) - \frac{\sigma^2}{\Delta x} & \text{if } i = 1\\
f_1(\theta_2) + \frac{\sigma^2}{\Delta x} & \text{if } i = 2
\end{cases} 
$$
And similarly for $\mathbf{A}^{\text{off}}$. Finally, from~\eqref{eq:Model:standard_control}, we define $\mathbf{A}^{\text{switch}}$ that describes the switch ON/OFF when a TCL exits the dead-band.  We define for indexing needs $ \delta := \bar{k} +1 - \underline{k}$. Notice that $ \rho_{i}(\tau) = \bar{\rho}(i, \tau)$, $ \rho_{i + \delta}(\tau) =\underline{\rho}(i, \tau)  ~\forall ~ \underline{k} \le i \le \bar{k}$, which means that a TCL turning off is described by a transition from the state $i$ to the state at the same temperature $i+\delta$, and vice-versa for a TCL turning on. This is better described by the schematic of the transitions in Figure~\ref{fig:TCL_diagram}. The nonzero elements of $\mathbf{A}^{\text{switch}}$ are then

\[ A^{\text{switch}}_{(i, \bar{k})} = 
\begin{cases} 
\frac{\sigma^2}{\Delta x} & \text{if } i = \bar{k}-1 \\
-f_1(\theta_{i+1}) -\frac{2\sigma^2}{\Delta x} & \text{if } i = \bar{k} \\
f_1(\theta_{i+1}) +\frac{\sigma^2}{\Delta x} & \text{if } i = \bar{k} + \delta + 1
\end{cases}
\]
\[ A^{\text{switch}}_{(i, \bar{k} +1)} = 
\begin{cases} 
- f_0(\theta_i) + \frac{\sigma^2}{\Delta x} & \text{if } i = \bar{k} - \delta -1 \\
f_0(\theta_{i}) -\frac{2\sigma^2}{\Delta x} & \text{if } i = \bar{k} +1 \\
\frac{\sigma^2}{\Delta x} & \text{if } i = \bar{k}+2.
\end{cases}
\].

{Assuming the heating power} $P_{h}$ large enough so that $ f_1(x) > 0\, \forall x \in [\theta_{--}; \theta_{++}]$, $\mathbf{A^{\text{nat}}}$ is a Metzler matrix (its off-diagonal elements are nonnegative) and $\one^\top\mathbf{A^{\text{nat}}}=\boldsymbol{0}^\top$. We now apply the {forward} Euler approximation {with discretization step $\Delta t$} to obtain the discrete time system
\begin{align}
    \label{eq:Model:linear_system_discrete}
    & \brho^{t+1} = \bPi^{\text{nat}}\brho^t & t = 0, ..., T
\end{align}
where $\bPi^{\text{nat}} = I+\Delta t \,\mathbf{A^{\text{nat}}}$. It is easy to see that $\bPi^{\text{nat}} \in \mc P^{N}$ (as defined in~\eqref{eq.transition.matrix.set}) if the diagonal elements of $\bPi^{\text{nat}}$ are non negative, which happens if the following condition is met:
\begin{equation}
\label{eq.discretization.condition}
 \max_{\theta \in [\theta_{--}, \theta_{++}]}\{|f_0(\theta) - \frac{2\sigma^2}{\Delta x}|, |f_1(\theta) + \frac{2\sigma ^2}{\Delta x}|\} \le \frac{\Delta x}{\Delta t}.
\end{equation}
The discretization previously proposed in \cite{paccagnan15}, that is, to approximate the probability density function with the average of the values at the extremes of the bins, guarantees the Metzler property (and, therefore, a valid discrete-time transition matrix) only when the system noise variance is assumed high enough. The opinion of the authors is that this assumption is not reasonable (for instance, it renders the model ill-defined in the nominal case with $\sigma = 0$). By contrast, condition \eqref{eq.discretization.condition} makes no assumption on the noise, it can be still satisfied in the nominal case and it is only related to the Euler approximation. It requires the discretization to be coarse enough in the temperature or fine enough in time, so that the model does not transfer more agents from a state than the ones actually present. 

Equation \eqref{eq:Model:linear_system_discrete} defines then the evolution of a Markov Chain where each state is {characterized} by the corresponding temperature bin and by the ON/OFF state. 
{Next we describe a typical control architecture for TCLs and we show that we can write the corresponding finite horizon optimal control problem as a problem of the form~\eqref{eq.opt.basic} satisfying Assumptions~\ref{ass.tractable} and~\ref{ass.feasible}.}

\subsection{TCL control as a convex optimization problem}

We now model the control action by assuming that a centralized controller can define the ON/OFF switching probabilities of a population of TCLs within the temperature dead-band. A possible realistic application setting in the context of electric grid optimization can be imagined in which the grid operator, after defining the control action, sends to a computer managing an aggregate of TCLs (which can be, for example, an apartment block)  the required probabilities of switching as a function of temperature bin and ON/OFF states. {Each TCL in the aggregate, after measuring its own temperature, can determine whether to switch ON or OFF according to the received required transition probabilities via an internal random number generator. Then, the aggregator manager polls the TCL states and estimates a probability distribution, which is sent back to the grid operator as a feedback. \par 
 {We denote by $\bPi^{\text{ctrl}}$ the transition probability matrix that takes into account the action of the controller.} For every state $\underline{k} \le j \le \bar{k}$, the controller can determine {freely} $\Pi^{\text{ctrl}}_{(j+\delta, j)} \in [0,1]$, which represents the probability of a TCL in state $j$ turning OFF, and the $\Pi^{\text{ctrl}}_{(j-\delta, j)} \in [0,1]$ with $ \bar{k}+1 \le j \le \bar{k} + \delta $, which represents the probability of a TCL in state $j$ turning ON. This control action will influence  other transition probabilities. {In particular,} let $\mc F \subset \Omega \times \Omega$ be the set of state {pairs} $(i,j)$ such that the transition probability from $i$ to $j$ cannot be modified directly by the controller. Assuming that, if no switch occur, the system evolves according to its natural transition probabilities and that no switch occur within the dead-band without a control action, we can write
\[
\Pi^{\text{ctrl}}_{(i,j)}  = (1- \Pi^{\text{ctrl}}_{(j\pm \delta, j)}) \Pi^{\text{nat}}_{(i,j)}~~ \forall ~ (i,j) \in \mc F
\]
where the sign $+$ or $-$ hold respectively when the system is required to switch OFF or ON {from state $j$}. {Then, a generic finite horizon optimal control problem for a population of TCLs takes the form}
{
\begin{subequations}
\label{eq.opt.tcls}
\begin{align}
\min_{\bPi^{t}\in\mc P^N,\brho^t\in\R^N_{\ge0}}~&~ f(\brho^1,\dots,\brho^T)\\
\text{s. to:}~&~\brho^{t+1} = \bPi^t \brho^t, ~~ t=0,\dots,T-1 \label{eq.opt.tcls.1}\\
~&~\Pi^{t}_{(i,j)}  = (1- \Pi^{t}_{(j\pm \delta, j)} )\Pi^{\text{nat}}_{(i,j)} ~ \forall ~ (i,j) \in \mc F \label{eq.opt.tcls.2}\\
~&~\mathbf{g}(\brho^{1},...,\brho^T) \le \boldsymbol{0} \label{eq.opt.tcls.3} \\
~&~\brho^0 = \bar\brho. \label{eq.opt.tcls.4}
\end{align}
\end{subequations}
}
{  The constraint \eqref{eq.opt.tcls.3} models additional constraints for the TCLs to the ones relative to the dynamics (for example, constraints on user comfort). If $f$ and $g$ are convex, then all constraints of~\eqref{eq.opt.tcls} satisfy Assumptions~\ref{ass.tractable} and~\ref{ass.feasible} and, using Proposition~\ref{propo.convex}, the change of variables~\eqref{eq.change.variables} leads to a convex reformulation of~\eqref{eq.opt.tcls}.}

\section{AC OPF problem with TCLs}



The goal of this section is to exploit the results of Proposition~\ref{propo.convex} to formulate (and solve) an AC optimal power flow (OPF) for power distribution systems featuring aggregations of TCLs as well as various DERs. 

Consider a distribution grid with $N+1$ nodes, with node $0$ representing the substation, modeled as an infinite bus. The nodes are represented by the set $\mathcal{N} \bigcup \{0\}$, $\mathcal{N} := \{1; \dots ; N\}$. Denote as $V_j \in \C$ the phasor of the voltage  at node $j$, respectively. For brevity, let $\mathbf{v} = [V_1, \dots , V_N] ^{\top} \in \C^N$, and $\mathbf{v}_{\text{mag}}^t := [|V_1|, \dots , |V_N|] ^{\top} \in \R^N$. Further, let $P_0^t$ denote the active power entering the substation.   Each node $j \in \mathcal{N} $ has uncontrollable  active and reactive loads, collected in the vector $\boldsymbol{\ell}^t_j \in \mathbb{R}^2$, where $t$ is the time index. Let $ \mathcal{G} := \{1, \ldots, G\}$ be the set of DER such as photovoltaic (PV) systems and energy storage systems. At time $t$ DER $i \in \mathcal{G}$ injects real and reactive powers, which are collected in the vector   $\mathbf{x}_i^t \in \mathbb{R}^2$. Per DER $i$ and time $t$, $\mathcal{X}_i^t \subset \R ^2$ is a set modeling hardware constraints; $\mathcal{X}_i^t$ is assumed to be convex and compact for all DERs. For example, for a PV system one has that 
$$
\mc{X}_i^t = \{ \mathbf{x}_i: 0 \leq x_{
i,1}\leq \bar{P}_{i}^{t}, x_{i,1}^2 + x_{i,2}^2\leq S_i^2 \}$$
where $\bar{P}_{i}^{t}$ is the maximum power generation (based on prevailing ambient conditions), and $S_i$ is the capacity of the inverter. 

It is well known that the non-linear 
AC power flow equations lead to a nonconvex formulation of the AC OPF problem. Here, we apply a linear approximations such as the ones proposed in, e.g.,~\cite{Bernstein18,Bernstein17,Bolognani15} (and pertinent references therein), to obtain an approximate linear relationships between voltage magnitudes and net injected powers of the form: 
\begin{align}
 \mathbf{v}_{\text{mag}}^t  \approx \sum_{k \in \cG} \mathbf{G}_k \mathbf{x}_k^t + \underbrace{\sum_{n \in \cN} - \mathbf{G}_n \boldsymbol{\ell}_{n}^t + \mathbf{a}}_{:= \bar{\mathbf{a}}^t} \label{eq:approx_v}
 \end{align}
 where the matrices $\mathbf{G}_k \in \R^{N \times 2}$ are built based on the location of the DERs and the non-controllable loads on the network, and $\mathbf{a} \in \R^{N}$ is a constant vector. Similarly, an approximate linear relationships between $P_0^t$ and net injected powers reads:
 \begin{align}
 P_{0}^t \approx \sum_{k \in \cG} \boldsymbol{\phi}_k^\top \mathbf{x}_k + \underbrace{\sum_{n \in \cN} - \boldsymbol{\phi}_n^\top \boldsymbol{\ell}_{n}^t + b}_{:= \bar{b}^t} \, .\label{eq:approx_p0} 
\end{align}
 With this model, a convex surrogate of the AC OPF can be formulated at each time $t$ as: 
\begin{subequations}
\label{eq:opf_problem_without_TCLs} 
\begin{align}
& \hspace{-.3cm} \min_{\{\mathbf{x}_i^t\}}           f(\{\mathbf{x}_i^t\})   \\
\text{subject to:}~&~ \mathbf{x}_k^t \in \mathcal{X}_k^t \, , \,\,\,\, \forall \,k  \in \mathcal{G} \\
& \mathbf{v}_{\text{min}}^t \leq \sum_{k \in \cG} \mathbf{G}_k \mathbf{x}_k^t   + \bar{\mathbf{a}} \leq \mathbf{v}_{\text{max}}^t \label{eq:opf_v} \\
& -\epsilon \leq P_{0,\text{ref}}^t -
\sum_{k \in \cG} \boldsymbol{\phi}_k^\top \mathbf{x}_k + \bar{b}^t \leq \epsilon \label{eq:opf_p0}
\end{align}
\end{subequations}
where $f(\{\mathbf{x}_i^t\})$ is a proper convex function modeling costs associated with the DERs;  $\mathbf{v}_{\text{min}}^t$ and $\mathbf{v}_{\text{max}}^t$ are vectors collecting minimum and maximum values for the voltage magnitudes, respectively; $P_{0,\text{ref}}^t$ is a target value for the power at the substation; and $\epsilon$ is a given tracking accuracy for the power at the substation.  

 Building on~\eqref{eq:opf_problem_without_TCLs}, consider now the case where the network has aggregations of (homogeneous) TCLs; let  $\mathcal{C}$ be the set of  aggregations. Assume that the $k$-th aggregation has $N_k$ states; let $\brho_k^t$ and $\mathbf{M}_k^t$ be quantities pertaining to the $k$-th aggregation of TCLs and refers to the time step $t$.      
 Defining the vector $ \mathbf{w}_k = [\mathbf{1}^{\top}_{N_k/2}, \mathbf{0}^{\top}_{N_k/2}]^{\top} $, one can write the expected number of TCLs in the $k$th aggregation to be in the ON state at time $t$ as $\mathbf{w}_k^{\top} \brho_k^t$. Therefore, 
the expected power consumed by the TCL aggregations $k$ at time $t$ reads: 
\begin{equation}
\label{eq:opf:TCL_power}
    P_{\text{TCL},k}(\brho_k^t) = \mathbf{w}_k^{\top} \brho_k^t P_{\text{TCL},k}^{\text{max}}  
\end{equation}
where $P_{\text{TCL},k}^{\text{max}}$ is the power consumption of the TCL aggregation when every agent is ON. 
Assuming no TCL reactive power consumption for simplicity, and therefore the vector of powers of the TCL $k$ is $[P_{\text{TCL},k}(\brho_k^t), 0]^\top$, one can modify~\eqref{eq:approx_v} and~\eqref{eq:approx_p0} to obtain an approximate expression for the expected voltage magnitudes and powers at the substation as:
\begin{align}
 & \mathbf{v}_{\text{mag}}^t  \approx \sum_{k \in \cG} \mathbf{G}_k \mathbf{x}_k^t - \sum_{j \in \cC} \mathbf{g}_j P_{\text{TCL},k}(\brho_k^t)  +  \bar{\mathbf{a}}^t \label{eq:approx_v_2}\\
 & P_{0}^t \approx \sum_{k \in \cG} \boldsymbol{\phi}_k^\top \mathbf{x}_k - \sum_{j \in \cC} \phi_{j,1} P_{\text{TCL},k}(\brho_k^t) +  \bar{b}^t \label{eq:approx_p0_2} 
\end{align}
where $\mathbf{g}_j$ relates $\mathbf{v}_{\text{mag}}^t $ and $P_{0}^t$ to the power consumed by the TCLs. Consider now formulating a multi-period OPF over $T$ time intervals; using notation and definitions above, \eqref{eq:opf_problem_without_TCLs} can be reformulated as follows to accommodate Problem \eqref{eq.opt.tcls} (opportunely reformulated applying Proposition \eqref{propo.convex}): 
\begin{subequations}
\label{eq:opf_problem_TCLs} 
\begin{align}
& \hspace{-.3cm} \min_{\substack{\{\mathbf{x}_i^t\},   
		  \{\brho_k^t\}
		  \{\bM_k^t \}}}           \sum_{t = 0}^T f^t(\mathbf{x}_i^t, \brho_k^t, \bM_k^t)   \\
& \hspace{-.3cm} \textrm{s.~to:~} \nonumber \\
& \mathbf{x}_k^t \in \mathcal{X}_k^t \, , \hspace{2.2cm} \forall \,k  \in \mathcal{G}, \forall ~ t = 0, \dots, T \\
& \mathbf{v}_{\text{min}}^t \leq \sum_{k \in \cG} \mathbf{G}_k \mathbf{x}_k^t   - \sum_{j \in \cC} \mathbf{g}_{j} P_{\text{TCL},k}(\brho_k^t) + \bar{\mathbf{a}} \leq \mathbf{v}_{\text{max}}^t \\
& -\epsilon  \leq  \sum_{k \in \cG} \boldsymbol{\phi}_k^\top \mathbf{x}_k - \sum_{j \in \cC} \phi_{j,1} P_{\text{TCL},k}(\brho_k^t) + \bar{b}^t - P_{0,\text{ref}}^t\leq \epsilon  \label{eq:opf_problem_TCLs:reference}\nonumber \\
& \hspace{3.7cm} \forall ~ t = 0, \dots, T  \\
& \mathbf{g}_k( \brho^0_k,..., \brho^T_k)\le \boldsymbol{0} \hspace{1.0cm} \forall \, k \in \mathcal{C} \\
& \bM^t_k\one_N = \brho^{t+1}_k \hspace{1.5cm} \forall \, k \in \mathcal{C}, \forall ~ t = 0, \dots, T-1  \\
& (\brho^t_k,\bM^t_k ) \in \mathcal{M}_k^{N_k}  
\hspace{1.1cm} \forall \, k \in \mathcal{C}, \forall ~ t = 0, \dots, T-1  \\
&  M^t_{k, (i,j)}  = (\rho_{k,j}^t- M^t_{k, (j\pm \delta,j)}) \Pi^{\text{nat}}_{k, (i,j)} \,\,\,\, \forall \, (i,j)\in\mc F_k^t, ~  \nonumber \\
& \hspace{3.3cm} \,\, \forall \, k \in \mathcal{C},\forall ~ t = 0, \dots, T-1  \\
& \brho^0_k = \bar{\brho}_k  \hspace{2.4cm} \forall \, k \in \mathcal{C}.
\end{align}
\end{subequations}
If the constraint $g_k( \brho^1_k,..., \brho^T_k)\le 0$ is convex, then problem~\eqref{eq:opf_problem_TCLs} is convex and can be efficiently solved using standard convex programming algorithms.

As a final remark, problem~\eqref{eq:opf_problem_TCLs} can be utilized as a building block for a model predictive control strategy, where the solution $\{\mathbf{x}^1_k\}, \{\brho^1_k, \mathbf{M}_k^1\}$ is implemented, and then~\eqref{eq:opf_problem_TCLs} is reformulated (and solved) once the window of $T$ time slots is advanced of one slot.

\section{Numerical experiments}

In this section, we illustrate the results of the paper on a modified version of the IEEE 37-node test feeder. We consider a single phase version of the feeder with real load and irradiance data measured in the Anatolia neighborhood (California) during a week in August 2012~\cite{Bank13}. 18 PV systems are located at nodes 4, 7, 10, 13, 17, 20, 22, 23, 26, 28, 29, 30, 31, 32, 33, 34, 35, 36. The ratings of these PV systems are
300 kVA for nodes 13, 17, 20, 22, 23, 26, 36, 350 kVA, 100 kVA for node 10 and 200 kVA for the remaining nodes. Let us denote the active power generated by the inverters of the PV systems by $P_{i,g}(t)$ and the reactive power generated as $Q_{i,g}(t)$. We denote as $S_i$ the rated complex power of the PV system $i$, which is a known constant, and $P_{i,g}^{\text{max}}(t)$ denotes the available maximum at each PV system. 
Three populations of 100 TCLs are located at nodes 8, 11 and 19. Each TCL consumes 4 kW for a total maximum consumption of 400 kW. The parameters of the TCL population were chosen among typical parameters for residential heat pumps \cite{Mathieu12} and they can be found in Table \ref{table:numerics:TCL_parameters}.
\begin{table}
    \centering
    \begin{tabular}{|c|c|c|c|c|c|c|c|c}
        \hline
         Param. & $C$& $R$ & $P$ & $\sigma$ & $\eta$ \\
         \hline
         Value & 1 & 2 & 4 & 0.001 & 3.5    \\
         Unit &kWh/$^\circ$C&$^\circ$C/kW & kW & $^\circ$C &- \\
         \hline
         Param. & $\theta_a$ & $\theta_+$ & $\theta_-$ & $\theta_{--}$ & $ \theta_{++} $ \\
         \hline
         Value  & 13 & 20 & 19 & 18 & 21 \\
         Unit & $^\circ$C & $^\circ$C & $^\circ$C  & $^\circ$C & $^\circ$C \\
         \hline
    \end{tabular}
    \caption{Parameters of the TCL model {for the stochastic differential equation~\eqref{eq:Model:tcl_ode}}}
    \vspace{-2em}
     \label{table:numerics:TCL_parameters}
\end{table}
The controller can command the PV inverters and alter the transition probabilities of the TCLs as discussed in the previous sections. The goals are (i) maintain the voltage at each bus between $0.95$ and $1.05$ power units and (ii) track a power reference signal at the substation.
 In order to simulate a sudden reduction in the power produced by the PV systems, (i.e. cloud coverage), the irradiance is reduced by 50\% at nodes 13, 17, 20, 22 and 23 around 12:00 and 14:00, as depicted in Figure \ref{fig:solar_power}.
 It is expected that the thermal energy stored in the TCLs can overcome this loss in solar power production. 
 The simulation is run over the timespan between 8 AM and 17 PM, with time discretization granularity $ \Delta_t = 20s $. {The TCLs are modeled with the procedure described in Section~\ref{sec.tcl_model} using a temperature discretization $\Delta x=0.1^\circ$C}. We choose an horizon length $T = 60$, which corresponds to a time of $20$ minutes. At every time-step the optimization {problem} \eqref{eq:opf_problem_TCLs} is solved { in a receding horizon fashion using a forecast of the load and irradiance}. {To guarantee feasibility, the constraint \eqref{eq:opf_problem_TCLs:reference} is substituted by a soft constraint introducing the slack variable $\epsilon$.}
We notice that the greater the norm of $\epsilon$ is, the greater the tracking error becomes. We therefore consider now $\epsilon$ as an optimization variable and we include in the cost a factor that penalizes its norm. We also note that \eqref{eq:opf_problem_TCLs:reference} implies $\epsilon\geq 0$. The cost function chosen is
\begin{align*}
 &f(\mathbf{x}_k^t, \bM_k^t) =  \frac{1}{T}\sum_{t=1}^t \Bigg[ \sum_{k\in \mc G}\Bigg( \gamma_P \frac{({P}_{k, g}^{\text{max}}(t)  - x_{k,1}^t)^2}{S_k^2} 
 \\ 
 &\hspace{1.3cm} + \gamma_Q\frac{(x_{k,2}^t)^2}{S_k^2} \Bigg) \!+\! \sum_{k \in \mc C} \sum_{(i,j)\notin \mc F_k^t}  \gamma_M M_{k, (i,j)}^t  + \gamma_{P_0} \epsilon \Bigg].
\end{align*}
The cost factor chosen were $\gamma_P = 3$ on the power curtailment of the PV systems, a cost $\gamma_Q = 2$ on the reactive power generation of the PV systems and a factor $\gamma_M = 1$ on the control action at the PV systems, aimed to penalize the probability of a switch happening, which is linked to premature deterioration of the system and discomfort of the user. A high enough weight of the soft constraint $\gamma_{P_0}$ guarantees that, if the problem \eqref{eq:opf_problem_TCLs} is feasible, then the soft constrained problem will have the same optimal solution (see e.g.,~\cite[Proposition 1]{bertsekas1975necessary}). Therefore, we choose $\gamma_{P_0} = 10^6$.
 The state evolution of each TCL is sampled from the transition probabilities determined by the controller. The grid is then simulated both using Matpower (nonlinear AC power flow) and the linearization~\eqref{eq:approx_v} and~\eqref{eq:approx_p0}.
 {Figure~\ref{fig:advantages} summarizes the advantages of having controllable TCLs in the system: the feeder is able to track the desired power reference at the substation even when the PV systems alone cannot. Furthermore, the total curtailment during the day is reduced, although it might happen that for some brief periods the controller increases the curtailment to take into account the user comfort and the constraints.  } 
{In Figures~\ref{fig:Voltages} and~\ref{fig:P_curtail}, we show the model mismatch between the linearized {\eqref{eq:approx_v} and~\eqref{eq:approx_p0}} and the nonlinear solution of the AC power flow equations computed by Matpower. While the voltages remain at an acceptable level, the power at the substation appears to have a constant offset. Future research will focus on off-set free control methods using either feedback-based optimization methods or adding integral action to the MPC controller.}
\begin{figure}
\input{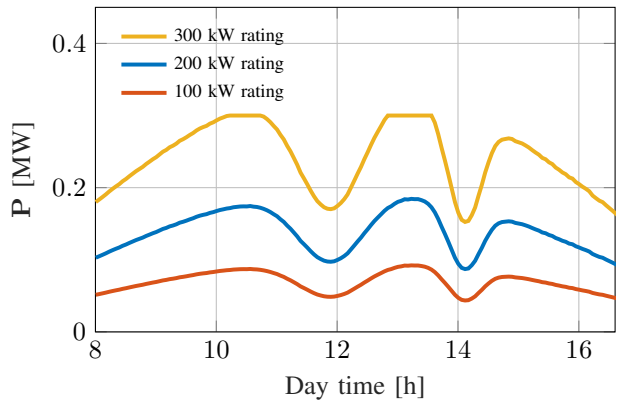}
  \caption{ Maximum instantaneous active power available at the PV systems.}
  \vspace{-2em}
  \label{fig:solar_power}
\end{figure}
\begin{figure}
\input{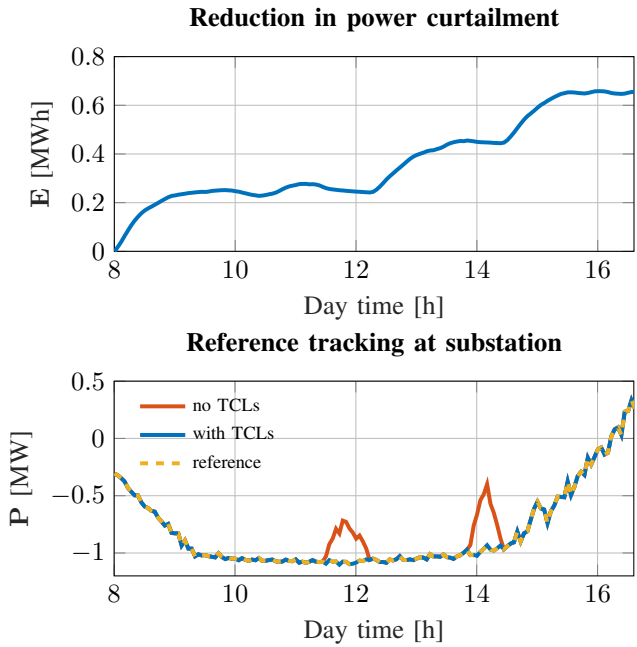}
  \caption{ Advantages in using the TCL control. {The first graph shows the cumulative difference in power curtailed by the PVs in one day. The second graph shows how this power is stored in the form of thermal energy by the TCLs and allows them to track the reference even when the PVs do not have enough instantaneous power.}}
  \label{fig:advantages}
\end{figure}
\begin{figure}

\input{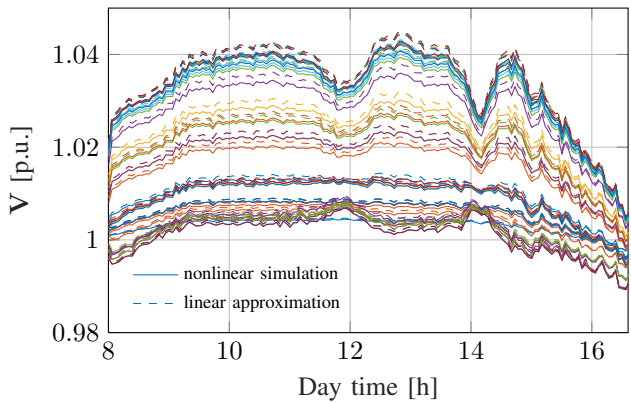}
  \caption{Comparison of linear and nonlinear simulation for the voltages at the buses. Despite the discrepancy there are no major voltage violations.}
  \label{fig:Voltages}
\end{figure}
\begin{figure}

\input{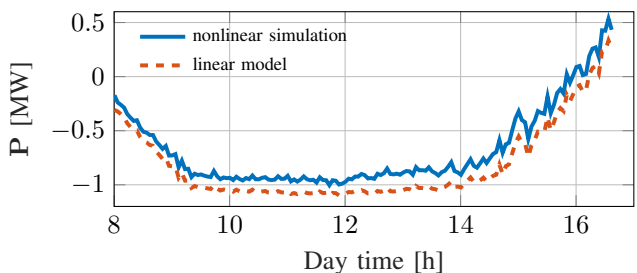}
  \caption{Comparison of the linear and nonlinear models for the power at the substation. The model mismatch creates a constant off-set. Future research will focus on off-set free control methods using either feedback-based optimization methods or adding integral action to the MPC controller.}
  \label{fig:P_curtail}
\end{figure}

\section{Conclusion}
In this paper we presented a method for solving a class of MDPs through an equivalent tractable convex programming problem. This class of problems can be a good representation of many problems concerning the control of a stochastic process in discrete time. To show a possible application, we devised a model of the evolution of a population of TCLs as a MDP and then we proposed a convex formulation of the OPF problem in the case when a bus is connected to a population of TCLs. Finally, a receding horizon control system was implemented and simulations were run to show the improvement on the performances of the grid when such a control is applied to the grid.

  \section*{Acknowledgments}
  The authors kindly acknowledge helpful discussions with Dario Paccagnan on modeling techniques for TCLs.

\bibliographystyle{IEEEtran}
\bibliography{bib/bib_file,bib/biblio_part2}

\end{document}